\documentclass{PoS}
\usepackage[american]{babel}
\usepackage{amsmath}
\usepackage{float}
\usepackage{cite}
\usepackage{enumerate}
\usepackage{slashed}
\def\b{\beta}

\newcommand{\beq}{\begin{equation}}
\newcommand{\eeq}{\end{equation}}
\newcommand{\bea}{\begin{eqnarray}}
\newcommand{\eea}{\end{eqnarray}}
\newcommand{\nn}{\nonumber}

\allowdisplaybreaks[1]

\title{AdS$_6$ T-duals and AdS$_6 \times S^2$ geometries in Type IIB}

\ShortTitle{AdS$_6$ T-duals and AdS$_6 \times S^2$ geometries in Type IIB}

\author{\speaker{Yolanda Lozano}$^a$, Niall T.~Macpherson$^{b}$
	and Jes\'us Montero$^{a}$\\ \\
	\llap{$^a$}Department of Physics, University of Oviedo,\\
	Avda. Federico Garc\'{\i}a Lorca 18, 33007 Oviedo, Spain.\\ \\
	\llap{$^b$}SISSA International School for Advanced Studies \\
	and INFN, sezione di Trieste, 34136 Trieste, Italy.\\ \\
	E-mail: \email{ylozano@uniovi.es}, \email{nmacpher@sissa.it},
	\email{jesus.montero.a@gmail.com}}

\abstract{We summarise the recent results in \cite{Lozano:2018pcp}, where the Type IIB AdS$_6$ solutions T-dual to the Brandhuber-Oz solution to massive Type IIA supergravity are shown to fit within an extension to allow for delocalized 5- and 7-branes of the global AdS$_6 \times S^2$ solutions constructed in \cite{DHoker:2017mds,DHoker:2017zwj}. We show that the Abelian T-dual solution provides an explicit example of a Riemann surface with the topology of an annulus.}
\FullConference{Corfu Summer Institute 2018 "School and Workshops on Elementary Particle Physics and Gravity"\\
		(CORFU2018)\\
		31 August - 28 September, 2018\\
		Corfu, Greece}

\begin{document}

\section{Introduction}

The AdS$_6/CFT_5$ correspondence remains much less understood than its counterparts in other dimensions. This is mainly due to the fact that in 5d there are no maximally supersymmetric CFTs. Indeed, in 5d the only superconformal algebra is $F(4)$, which contains 8 Poincar\'{e} and 8 conformal supercharges for a total of 16 superconformal charges\cite{Nahm:1977tg}.
Accordingly, AdS$_6$ backgrounds are quite unique: The only solutions to massive Type IIA supergravity are the Brandhuber-Oz (BO) background  \cite{Brandhuber:1999np}, and orbifolds thereof \cite{Bergman:2012kr}, which contain singularities associated to the presence of an O8 orientifold fixed plane. 

AdS$_6$ solutions should provide interesting backgrounds in which to realise holographically five dimensional CFTs. Even if gauge theories in 5d are non-renormalizable, string theory arguments show that, for specific gauge groups and matter content, they can flow to strongly coupled CFTs in the UV  \cite{Seiberg:1996bd,Intriligator:1997pq,Jefferson:2017ahm}. The holographic study of such fixed point theories is especially relevant, since these theories do not allow a standard Lagrangian description. 
%This direction of study remained however largely unexplored due to the scarcity of explicit AdS$_6$ solutions in supergravity.

%For these gauge theories, string theory realisations are known in some cases that can be used to construct their fixed point AdS$_6$ duals. 
In particular, the Brandhuber-Oz background arises as the near-horizon geometry of a stack of $N$ D4-branes probing an orientifold 8-plane with $N_f$ D8-branes. The holographic dual is the UV fixed point of a 5d gauge theory with $Sp(N)$ gauge group, an antisymmetric hypermultiplet and $N_f<8$ fundamental hypermultiplets \cite{Seiberg:1996bd}.
More general string theory realisations of 5d fixed point theories can be given in terms of $(p,q)$ 5-brane webs in Type IIB \cite{Aharony:1997ju,Aharony:1997bh}. In these set-ups the 5-branes share 4 directions, and each 5-brane is represented with a line (whose slope is related to the type of brane) on the plane where the web is realised.
% realisations of Hanany-Witten brane set-ups.
The corresponding 5d field theory lives on D5-branes which are of finite extent on this web plane. Flavours are realised in simple cases through external semi-infinite D5-branes, and by adding to the web optional 7-branes \cite{DeWolfe:1999hj}. The corresponding supergravity duals should describe the origin of the Coulomb branch of these theories, in the limit in which the 5d gauge coupling is sent to infinity. In the brane set-up, this is equivalent to taking the internal D5-branes to be coincident and of zero length in the plane. One obtains in this way a collapsed 5-brane web, from which only the external 5-branes remain.

However, the strong constraints imposed by supersymmetry suggested that no AdS$_6$ solutions existed in Type IIB \cite{Passias:2012vp}. Indeed, the Type IIB AdS$_6$ duals to the fixed point theories originated from $(p,q)$ 5-brane webs remained unknown until the recent work of  D'Hoker, Gutperle and Uhlemann (DGU) \cite{DHoker:2016ysh,DHoker:2017mds,DHoker:2017zwj}. These  authors constructed global physically sensible backgrounds for the local solutions previously found in \cite{D'Hoker:2016rdq} (DGKU). These DGU geometries take the form of a warped product of an AdS$_6\times S^2$ geometry over a Riemann surface $\Sigma$, which needs to have a boundary for regularity. This boundary contains poles that are associated to semi-infinite $(p,q)$ 5-branes. The addition of 7-branes is realised through branch points in $\Sigma$ which introduce non-trivial monodromies along the corresponding branch cuts.
A number of works support the holographic interpretation of these global solutions as gravity duals to 5d CFTs living in $(p,q)$ 5-brane webs, see e.g.~\cite{Gutperle:2017tjo,Kaidi:2017bmd,Gutperle:2018wuk,Fluder:2018chf}. Some of these tests have been extended to include 7-branes  \cite{Gutperle:2018vdd,Bergman:2018hin,Chaney:2018gjc}. Additional support for the holographic interpretation of the 5-brane solutions in \cite{DHoker:2017mds} has been provided very recently by uplifting the $(p,q)$ 5-brane webs to punctured M5-branes wrapping a Riemann surface \cite{Kaidi:2018zkx}.

T-duality played a very important role in motivating these developments. Indeed, the first AdS$_6$ solutions to Type IIB supergravity were constructed in \cite{Lozano:2012au,Lozano:2013oma}  acting with Abelian and non-Abelian T-duality \cite{Sfetsos:2010uq} (ATD and NATD, respectively) on the BO solution. This raised the interest in the study of classifications of AdS$_6$ solutions \cite{Apruzzi:2014qva,Kim:2015hya,Kim:2016rhs,Gutowski:2017edr}, which culminated with the work of DGKU \cite{D'Hoker:2016rdq} and DGU \cite{DHoker:2016ysh,DHoker:2017mds,DHoker:2017zwj}. The explicit realisation of the T-dual backgrounds as DGU solutions has recently been worked out in \cite{Lozano:2018pcp}, whose main results are summarised in this review article.

The Abelian T-dual solution describes the same 5d $Sp(N)$ fixed point theory as the Brandhuber-Oz solution \cite{Bergman:2012kr}, but now in terms of a system of smeared D5, NS5 and D7/O7 branes in Type IIB. However, the CFT dual of the non-Abelian T-dual solution is not known. In principle, it may be different from the previous fixed point theory, because NATD has not been proved to be a string theory symmetry. In fact, in several examples the non-Abelian T-dual background has been argued to be associated to QFTs living on (Dp,NS5) Hanany-Witten \cite{Hanany:1996ie} brane set-ups, realising a different CFT than the original gravity solution \cite{Lozano:2016kum,Lozano:2016wrs,Lozano:2017ole,Itsios:2017cew}.

This review article is organised as follows. We first provide a brief introduction to the solutions of DGKU/DGU in section \ref{sec:alaDHoker}. Next, we describe in section \ref{sec:Tduals} how the Abelian T-dual and non-Abelian T-dual solutions of the BO background fit globally within an extension of  the formalism of DGU. In section \ref{sec:field-theory} a possible interpretation of the CFT dual to the non-Abelian T-dual solution is discussed. We end with a summary and discussion of these results in section \ref{sec:conclusions}.

\section{The DGKU/DGU  AdS$_6\times S^2$ solutions to Type IIB}\label{sec:alaDHoker}

In this section we summarise the main features of the AdS$_6\times S^2\times \Sigma$ solutions to Type IIB supergravity constructed in \cite{D'Hoker:2016rdq} and \cite{DHoker:2017mds}, including 7-branes \cite{DHoker:2017zwj}. The solutions constructed in \cite{DHoker:2017mds} take the form of a warped product of an AdS$_6\times S^2$ geometry over a Riemann surface $\Sigma$, which needs at least one boundary for regularity. Information about the dual 5d SCFT is encoded through poles lying on the boundary, that are associated to external branes in the 5-brane web. The addition of 7-branes is implemented allowing for punctures (branch points) in $\Sigma$ which introduce non-trivial $SL(2,\mathbb{R})$ monodromies. The regularity conditions provided guarantee regularity in 10d away form localised sources and their associated branch cuts. Close to these singular points, the supergravity fields exhibit the expected near brane behavior, including the monodromies induced by turning around the 7-branes. 

In \cite{D'Hoker:2016rdq} the problem of finding supersymmetric AdS$_6$ solutions to type IIB supergravity was reduced to finding two locally holomorphic functions $\mathcal{A}_{\pm}$ \footnote{This formalism is reminiscent of the half BPS AdS$_4$ solutions constructed in \cite{D'Hoker:2007xy,D'Hoker:2007xz}. An alternative formulation with AdS$_6$ geometries in one to one correspondence with the solutions of a single Laplace equation and its derivatives was derived in \cite{Macpherson:2016xwk,Apruzzi:2018cvq}.}. The local form of the AdS$_6$ solutions in \cite{D'Hoker:2016rdq} is as follows
\begin{align}\label{eq:metric_DHoker}
ds^2 &= \lambda_6^2 ds^2(AdS_6)+ \lambda_2^2 ds^2(S^2)+ds^2(\Sigma),\quad ds^2(\Sigma)= 4 \tilde{\rho}^2dw d\overline{w} ,\nn\\[2mm]
B_2+i\, C_2 &= \mathcal{C}\wedge \text{Vol}(S^2),
%  H_3+ i F_3 &= d\mathcal{C}\wedge \text{Vol}(S^2),
\end{align}
where, $\lambda_6^2,\lambda_2^2, {\tilde \rho}^2, \mathcal{C}$, the dilaton and axion are functions defined on $\Sigma$, and can be expressed in terms of the following:
\begin{align}\label{eq:functions_DHoker}
&\kappa_{\pm}= \partial_w \mathcal{A}_{\pm},\quad \kappa^2= -|\kappa_+|^2+ |\kappa_-|^2,\quad \partial_w \mathcal{B} = \mathcal{A}_+\partial_w\mathcal{A}_--\mathcal{A}_-\partial_w\mathcal{A}_+,\\[3mm] 
&\mathcal{G}= |\mathcal{A}_+|^2-|\mathcal{A}_-|^2+\mathcal{B}+\overline{\mathcal{B}},\quad W = R+\frac{1}{R} = 2 + 6 \frac{\kappa^2\mathcal{G}}{|\partial_w\mathcal{G}|^2},~~ \kappa^2= -\partial_w \partial_{\overline{w}}\mathcal{G}\nonumber \,,
\end{align}
which in turn depend on the two locally holomorphic functions $\mathcal{A}_{\pm}$ determining the solution. The warp factors of the string frame metric are then given by\footnote{We use the usual definition for the dilaton $\Phi$, which is related to that appearing in \cite{D'Hoker:2016rdq} as $\Phi=2\phi$. Note also that Einstein frame metric is used in that paper.}:
\beq\label{eq:factors_DHoker}
\lambda_2^2 =e^{\Phi}\frac{ c^2 \kappa^2(1-R)}{9 \tilde{\rho}^2(1+R)},\quad \lambda_6^2= e^{\Phi}\frac{c^2 \kappa^2 (1+R)}{\tilde{\rho}^2(1-R)},~~~\tilde{\rho}^2=  e^{\Phi/2}\frac{\sqrt{R+R^2}}{|\partial_w \mathcal{G}|}\left(\frac{\kappa^2}{1-R}\right)^{3/2}.
\eeq
The axion-dilaton field $\tau = C_0+ i e^{-\Phi}$ and $F_1=dC_0$ are derived from
%\beq\label{eq:axio-dilaton_DHoker_1}
%B=\frac{1+i \tau}{1-i \tau},~~~\tau = C_0+ i e^{-\Phi},\qquad F_1= dC_0,
%\eeq
\beq\label{eq:axio-dilaton_DHoker_2}
B= \frac{1+i \tau}{1-i \tau} = \frac{\kappa_+ \partial_{\overline {w}} \mathcal{G}- \overline{\kappa}_- R \partial_w \mathcal{G}}{\overline{\kappa}_+ R  \partial_w \mathcal{G}- \kappa_-  \partial_{\overline {w}} \mathcal{G}}\, .
\eeq
Finally, a complex function
\beq\label{eq:C_potential_DHoker}
\mathcal{C} = \frac{4 i}{9} \bigg(\frac{\overline{\kappa}_- W\partial_w \mathcal{G}-2 \kappa_+ \partial_{\overline{w}}\mathcal{G}}{(W+2)\kappa^2}-\overline{\mathcal{A}}_--2\mathcal{A}_+-\mathcal{K}_0\bigg),
\eeq
where $\mathcal{K}_0$ is an integration constant, provides the 2-form NS and RR potentials:
\[ {\cal C}\,\text{Vol}(S^2)= B_2+i\ C_2 \,. \]

As expected for the gravity duals of 5d SCFTs with $F(4)$ superalgebra, these supergravity backgrounds preserve 16 supersymmetries, and are invariant under the $SO(2,5) \oplus SO(3)$ maximal bosonic subalgebra of $F(4)$.

We next summarise the global solutions found in \cite{DHoker:2017mds,DHoker:2017zwj}. We review the formulation when $\Sigma$ is the upper half-plane, relevant for the study of the non-Abelian T-dual solution, and when $\Sigma$ is the annulus, which is associated to the Abelian T-dual solution. In this latter case the formalism in \cite{DHoker:2017mds} needs to be extended to include 7-branes in the annulus. The reader is referred to  \cite{Lozano:2018pcp} where this extension was worked out in detail.

%We reproduce the non-Abelian T-dual solution as a DGU geometry when $\Sigma$ is the upper half-plane, and the Abelian T-dual when $\Sigma$ is an annulus. We review both formulations introduced by DGU below. We also generalise the annulus construction in \cite{DHoker:2017mds} to include 7-branes. 

We focus on solutions with D7-branes, as this is enough to reproduce the T-duals. Geometries with monodromies associated with general $(p,q)$ 7-branes can be obtained for the upper-half plane considering the action of $SL(2,\mathbb{R})$ on the locally holomorphic functions $\mathcal{A}_\pm$, as detailed in \cite{DHoker:2017zwj}. However, it is not clear that $(p,q)$ 7-branes other than D7's ($p=\pm1$)  can be accommodated on the annulus \cite{Lozano:2018pcp}.

\subsection{Global solutions for the upper half plane}
\label{sec:upper}

Following \cite{DHoker:2017zwj}, a solution arising from $L$ semi-infinite 5-branes and D7-branes is associated to the following functions:
\begin{equation}
\label{Apm7branes}
{\cal A}_\pm = {\cal A}_\pm^0+\sum_{l=1}^L Y_\pm^l\log{(w-p_l)} +\eta\int_\infty^w dz\, f(z) \sum_{l=1}^L \frac{Y^l}{z-p_l},
\end{equation}
where $p_l$ with $l=1,\ldots,L$ are the poles where the 5-branes are located, $Y^l= Z^l_+-Z^l_-$, $Y_\pm^l=\eta\, Z_+^l$, $Z_+^l= - \overline{Z_-^l}= \frac{3}{4}(q-ip)$ gives the charge of the corresponding $(p,q)$ 5-brane in the absence of D7-branes\footnote{The charges of the external 5-branes are associated with the residues of $\partial_w \mathcal{A}_\pm$ at the poles $p_l$, and are given by $\mathcal{Y}_\pm^l= Y_\pm^l + \eta f(p_l) Y_l$. In the absence of 7-branes, $f(z)\equiv 0$ and these are reduced to $\mathcal{Y}_\pm^l=Z_\pm^l$.}, $\eta=+1$ for D7-branes and $\eta=-1$ for anti-D7's, and ${\cal A}_\pm^0$ are constants.
The integration contour has been chosen such that it avoids crossing the poles on the boundary, as well as the punctures and corresponding branch cuts in $\Sigma$. The $f(w)$ function encodes the information about the branch cut structure of the punctures associated to D7-branes, 
\begin{equation}
\label{fomega}
f(w)=\sum_{i=1}^I \frac{n_i ^2}{4\pi}\log{\Bigl(\gamma_i \frac{w-w_i}{w-\bar{w}_i}\Bigr)},
\end{equation}
where $w_i$ is the locus of a D7-brane puncture, $\gamma_i$ is a phase specifying the orientation of the associated branch cut\footnote{Such that $\gamma_i=+1$  gives a branch cut extending in the negative imaginary direction and the opposite for $\gamma_i=-1$.},
and $n_i^2$ with $i=1,\ldots,I$ is the number of D7-branes at the $i$th-puncture. 
%A solution with no D7-branes (and therefore no monodromy) can be recovered by setting all $n_i^2=0$.

For backgrounds generated from the locally holomorphic functions given in eq.~\eqref{Apm7branes}, a series of conditions must be satisfied for the solutions to be regular\footnote{Most of these conditions derive from the requirement that the functions $\mathcal{G}$ and $\kappa^2$ are both positive or negative in the interior of $\Sigma$ and vanish on the boundary.}. These reduce the number of free parameters, relating for instance the charges of the external 5-branes encoded in $Z_\pm$ to the positions of the poles and the integration constants ${\cal A}_\pm^0$. See \cite{DHoker:2017zwj} or \cite{Lozano:2018pcp} for more details.

The non-Abelian T-dual of the Brandhuber-Oz solution discussed in section \ref{sec:NATD} contains  smeared 5- and 7-branes. Given that the associated singularities are not isolated points as in the DGU solutions, this background is not expected to satisfy the aforementioned regularity conditions.

\subsection{Global solutions for the annulus}

The formalism of DGU \cite{DHoker:2017mds} also describes 5-brane solutions when $\Sigma$ is an annulus, defined as the set of points satisfying $0\leq {\rm Re}(w)\leq 1$ and $0\leq {\rm Im}(w)\leq t/2$, with periodicity under $w\rightarrow w+1$ and $t>0$. In \cite{Lozano:2018pcp} an extension of this formalism  to include 7-branes was presented. The corresponding locally holomorphic functions are given by:
\beq\label{eq:annulusD7}
{\cal A}_{\pm} ={\cal  A}_{\pm}^0 +\sum_{l=1}^L Y^l_{\pm} \log(\theta_1(w-p_l|\tau))+\eta\int_1^w dz f(z)\sum_{l=1}^L Y^l\partial_{z}\log(\theta_1(z-p_l|\tau)) \,,
\eeq
where it is assumed that the reference point $w=1$ is regular, $Y^l$, $Y^l_{\pm}$, $\eta=\pm1$ were introduced below \eqref{Apm7branes}, and $\theta_1(w|\tau)$ are the Jacobi theta functions of the first kind with modular parameter $\tau=i\,t$,
\beq
\theta_1(w|\tau)= 2 \sum_{n=0}^{\infty} (-1)^n e^{i\pi(n+\frac{1}{2})^2\tau}\sin((2n+1)\pi w),
\eeq
which satisfy the following periodicity properties:
\begin{eqnarray}
&&\theta_1(w+1|\tau)=-\theta_1(w|\tau)\,,\nonumber\\ 
&&\theta_1(w+\tau|\tau)=-\theta_1(w|\tau)\exp{(-i\pi\tau-2\pi iw)}\,.
\end{eqnarray}
The function $f(w)$ describing the monodromies of D7-branes takes the following form for the annulus \cite{Lozano:2018pcp}:
\beq
\label{f-annulus}
f(w) = \sum_{i=1}^I\frac{ n_i^2}{4 \pi}\bigg(\log\left(\gamma_i \frac{\theta_1(w-w_i|\tau)}{\theta_1(w-\overline{w_i}|\tau)}\right)- \frac{2\pi i }{\tau} (w_i-\overline{w_i})w\bigg).
\eeq 
Both the use of theta functions with the aforementioned periodicity properties and the particular form of $f(w)$ in \eqref{f-annulus} are required for the periodicity of the geometry under $w \rightarrow w+1$. They are also required for the 10d solution to be well-defined and regular (away from the poles) at both boundary components of $\Sigma$, $\mathbb{R}$ and $\mathbb{R}+\frac{\tau}{2}$. We remark, however, that regularity and periodicity are only achieved when also a number of supplementary  conditions hold, as detailed in \cite{Lozano:2018pcp}. These conditions cannot be enforced on a background arising from smeared branes, like the Abelian T-dual of the BO solution.

%\beq\label{eq:acondand7}
%{\cal A}^0_{\pm}=-\overline{{\cal A}^0_{\mp}}.
%\eeq

\section{The T-duals of Brandhuber-Oz as DGU solutions}
\label{sec:Tduals}

\subsection{The Brandhuber-Oz AdS$_6$ solution}
\label{sec:BO}

The Brandhuber-Oz background arises as the near-horizon geometry of a stack of $N$ D4-branes probing a system of an O8-plane with $N_f$ D8-branes. It can be expressed as a warped product of AdS$_6$ times half a 4-sphere,
\vspace{-1mm}
\begin{align}\label{eq:AdS6 sol}
ds^2&= \frac{L^2 W^2}{4}\bigg[9 ds^2(AdS_6)+ 4\bigg(d\theta^2+\sin^2\theta ds^2(S^3)\bigg)\bigg],\qquad e^{\Phi}=\frac{2}{3L} W^{5},\\[2mm]
F_0&=m,\quad F_4 = \frac{5L^4}{W^2}\sin^3\theta d\theta\wedge \text{Vol}(S^3),\quad W= (m \cos\theta)^{-\frac{1}{6}},\nn
\end{align}
where $L$ denotes the AdS$_6$ radius and $\theta\in [0,\frac{\pi}{2}]$. The orientifold action is realised as $\theta\rightarrow \pi-\theta$, with the O8-fixed plane located at $\theta=\pi/2$, where the $S^3$ boundary is singular. The geometry covers half of the $S^4$, and therefore only one side of the orientifold fixed plane. 
The length scale $L$ and the Romans mass $m$ are related to the quantised charges, which are computed from the RR fluxes,
\beq
N =\frac{1}{2\kappa_{10}^2T_4}\int_{S^4}F_4= \frac{9}{16\pi }L^4m^{1/3}, \qquad N_{8}=\frac{m}{2\kappa_{10}^2T_8}=2\pi \,m,
\eeq
where $ 2\kappa_{10}^2T_p= (2\pi)^{7-p}$, $N_{8}\equiv 8-N_f$ accounts for the charge of the O8 plane and the $N_f$ D8-branes, $N$ and $N_8$ must be integers and $N_f \leq 7$. The supergravity limit is valid when
\beq
N_{8}^{\;3}\, N>>1,~~~~ \frac{N}{N_{8}}>>1\,.
\eeq

This solution preserves 16 supersymmetries and exhibits the explicit $SO(2,5)$ conformal symmetry and the $SU(2)_R$ R-symmetry of the bosonic subalgebra of $F(4)$, plus a global $SU(2)$ symmetry\footnote{The would-be $SO(5)$ global symmetry of the full 4-sphere is broken down to $SO(4)\sim SU(2)_R \times SU(2)$ for the half $S^4$ of the solution.} associated to the antisymmetric hypermultiplet of the dual $Sp(N)$ theory.

\subsection{The AdS$_6$ Abelian T-dual solution}
\label{sec:ATD}

Applying Abelian T-duality on the Hopf fibre of the $S^3$ factor in \eqref{eq:AdS6 sol}, the following supergravity solution is found \cite{Lozano:2012au,Lozano:2013oma}:
\begin{align}\label{eq:AdS6_sol_ATD}
ds^2&= \frac{L^2 W^2}{4}\bigg[9 ds^2(AdS_6)+ \sin^2\theta ds^2(S^2)\bigg]+\frac{4}{L^2W^2\sin^2\theta}\left(d\psi^2+\frac{1}{4}L^4W^4\sin^2\theta d\theta^2\right) ,\nn\\[2mm]
e^{-\Phi}&=\frac{3L^2}{4W^4}\sin\theta,~~~B_2=\cos{\xi}d\psi\wedge d\phi,  
~~~F_1=-m d\psi,\\[2mm] 
F_3&=\frac{5 L^4}{8 W^2}\sin^3\theta d\theta\wedge\text{Vol}(S^2),~~~~~ F_5=0.\nn
\end{align}
The Hopf fibre becomes the T-dual direction $\psi \in [0,\pi]$, and  $(\xi,\phi)$ parametrise the $S^2$  associated to the $SU(2)$ R-symmetry of $F(4)$, such that $\text{Vol}(S^2)=\sin{\xi}d\xi \wedge d\phi$. 

This background arises in the near-horizon limit of a system of D5, NS5 and D7/O7 branes, and inherits the singularity at $\theta=\frac{\pi}{2}$, where the D8/O8 system in IIA is located, which now corresponds to a D7/O7 system. Additionally, T-dualising on a circle (or 3-sphere) of shrinking size gives rise to a singularity. In our case, this implies a second singularity placed at $\theta=0$, which is associated to smeared NS5-branes sourcing the newly generated $B_2$ field. The presence of the D7/O7 and NS5-branes can also be inferred from the behavior of the supergravity fields in \eqref{eq:AdS6_sol_ATD} close to the singularities. %\footnote{Notwithstanding, the presence of the orientifold fixed plane in the D7/O7 system cannot be told directly from the supergravity fields. Instead, one needs to consider the transformation of the orientifold action under T-duality, see appendix B of \cite{Lozano:2018pcp} for details.}.

Again, we can relate the parameters of the solution to the quantised charges (computed as Page charges in the presence of a $B_2$ field),
\begin{align}\label{eq:Page_charges_ATD}
N_{7}&=\frac{1}{2\kappa_{10}^2 T_7}\int_{\psi=0}^{\psi=\pi} F_1=-m\pi ,\nn\\[2mm]
%N_{D5}^r&=\frac{1}{2\kappa_{10}^2 T_5}\int_{S^2}\int_{r=0}^{r=\pi} (F_3-B_2\wedge F_1) = -\frac{1}{2} N_{7},\nn\\[2mm]
N_{D5}&=\frac{1}{2\kappa_{10}^2 T_5}\int_{S^2}\int_{\theta=0}^{\theta=\frac{\pi}{2}} (F_3-B_2\wedge F_1)=\frac{N}{2} =\frac{9}{32\pi }L^4m^{1/3}   ,\nn\\[2mm]
N_{NS5}&=\frac{1}{2\kappa_{10}^2 T_5}\int_{S^2}\int_{\psi=0}^{\psi=\pi} H_3 =1
\end{align}
%\red{(Anything on the sign discrepancy for $N_7$ between \eqref{eq:ATD_charges} and \eqref{eq:Page_charges_ATD}?)\\}
which are all integers. Note that the T-duality transformation maps the O8 plane onto two O7 planes with charge 4, and the $N_f$ D8-branes onto $N_f/2$ D7-branes (plus their mirrors). $N_7$ accounts for the charge of each D7/O7 system. NS5-branes are created as well at the point where the original $S^1$ shrinks. In terms of these charges, the supergravity limit is valid when 
\beq
N_{7}\, N_{D5}>>1,~~~\frac{N_{D5}}{N_{7}}>>1.
\eeq

The Abelian T-dual solution describes holographically the 5d $Sp(N)$ SCFT dual to the Brandhuber-Oz solution \eqref{eq:AdS6 sol} in terms of smeared D5, NS5 and D7/O7 branes in Type IIB. All the supersymmetries of the original BO background are preserved by T-duality at the supergravity level. Still, the global $SU(2)$ symmetry associated to the antisymmetric hypermultiplet is reduced to $U(1)$\footnote{This symmetry is however enhanced to $SU(2)$, in agreement with the enhanced mesonic symmetry \cite{Bergman:2012kr}.}.

\subsubsection{The AdS$_6$ Abelian T-dual solution as a DGU geometry}

\begin{figure}
	\centering
	\includegraphics[scale=1.3]{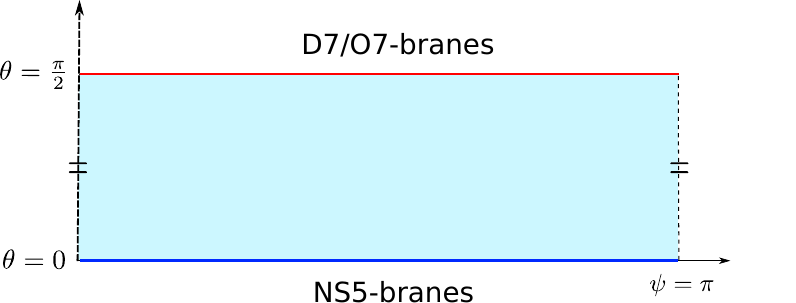}
	\caption{The annulus for the Abelian T-dual background. NS5-branes are smeared along the lower boundary at $\theta=0$, and 
		D7/O7 branes are smeared at the upper boundary at $\theta=\pi/2$. The annulus topology follows from the periodicity under $\psi\rightarrow \psi+\pi$, or $w\rightarrow w+1$, shifts.}
	\label{fig:annulus_ATD}
\end{figure}

In order to embed the Abelian T-dual into the global solutions classified by DGU, we first need to identify the Riemann surface $\Sigma$. Given the periodicity of the T-dual direction $\psi$, $\Sigma $ has the topology of an annulus. Parametrising it by the $(\psi,\theta)$-directions of \eqref{eq:AdS6_sol_ATD}, we can choose the local holomorphic coordinate $w$ to be\footnote{With this choice the periodicity under $\psi\rightarrow \psi+\pi$ translates into the identification $w\rightarrow w+1$ of the annulus, and NS5-branes are smeared along the real line. This is different from the choice in \cite{D'Hoker:2016rdq}, where $w = (\cos\theta)^{2/3}+ i \beta \psi$.} 
\begin{equation}
\label{omega}
w=\frac{1}{\pi}\Bigl(\psi+\frac{i}{\beta}(1-(\cos{\theta})^{2/3})\Bigr), 
\end{equation}
where $\beta=4 m^{1/3}/(3 L^2)$, so that in the supergravity limit $\beta\rightarrow 0$. The resulting annulus is depicted in Figure \ref{fig:annulus_ATD}. 
With this choice for the $w$ coordinate, the locally holomorphic functions giving rise to the supergravity fields of the T-dual solution in \eqref{eq:AdS6_sol_ATD} are\footnote{These ${\cal A}_\pm$ functions provide a \textit{local} embedding for the T-dual solution, which was already presented in \cite{D'Hoker:2016rdq}.}
\begin{equation}
\label{eq:holo_Ansatz_ATD_new}
{\cal A}_\pm =\frac{3}{8}m\pi^2 \Bigl(w^2-\frac{2i}{\beta\pi}w-\frac{1}{(\beta\pi)^2}\Bigr)
\pm i\frac34 \pi \Bigl(w-\frac{i}{\beta\pi}\Bigr) .
\end{equation}

The Abelian T-dual solution contains smeared NS5-branes at $\theta=0$ and smeared D7/O7 branes close to $\theta=\pi/2$. We therefore consider poles associated to NS5-branes at ${\rm Im}(w)=0$ and punctures associated to D7/O7 branes at ${\rm Im}(w)=\frac{1}{\beta\pi}$, which lie at infinity in the supergravity limit $\b \rightarrow 0$. We take these NS5 and D7/O7 branes to account for the charges of the external branes. Considering this limit, and the smearing of the branes, the above ${\cal A}_\pm$ can be reproduced. 

Let us start by analysing the contribution of the NS5-branes, given by the term with $Y^l_{\pm}$ coefficients in \eqref{eq:annulusD7}. We first note that the position of the upper boundary at ${\rm Im}(w)=1/(\beta\pi)$ compared to the definition of the annulus $0\leq{\rm Im}(w)\leq t/2$ leads to $t= 2/(\beta\pi)$, that grows large in the $\b\rightarrow 0$ limit.
Given that $\tau=i\,t$, this allows us to use the asymptotic expansion for the Jacobi theta-function,
\begin{equation}
\theta_1(z\,|\,\tau) \Big\lvert_{t\to\infty}=2 e^{-\frac{\pi}{4} t}\sin(\pi z)+...
\end{equation}
Besides, we model the smearing of the NS5-branes computing the charge as:
\[\sum_{l=1}^L Y_\pm^l=\mp\frac34 \frac{1}{\pi}\int_0^\pi d\psi=\mp\frac34\,.\]
Putting together the above expansion of the theta functions for $\b\rightarrow 0$ and the smearing of the branes, we arrive at:
\begin{equation}
{\cal A}_\pm \Big\lvert_{NS5}=\pm \,i \frac34 \pi \,w \,,
\end{equation}
which exactly reproduces the linear term in $w$ not proportional to the mass (and therefore arising from NS5-branes) in equation (\ref{eq:holo_Ansatz_ATD_new}). Here we have considered that, according to the conventions in eq.~\eqref{eq:Page_charges_ATD}, we have anti D7-branes and therefore $\eta=-1$ and $Y^l_\pm=-Z^l_\pm$. Remark that we have omitted constant terms, which can absorbed in ${\cal A}^0_\pm$. 

Let us now consider the ${\cal A}_\pm |_{D7/O7}$ contribution of the D7/O7 punctures, given by the integral term in \eqref{eq:annulusD7}, with $f(z)$ as in \eqref{f-annulus}. We use the same expansion for the theta functions as above, and take into account that the D7/O7 system lies at $z_i=\psi/\pi +i/(\beta\pi)$, i.e.~at $\theta=\pi/2$ and smeared on the $\psi$-direction. The sum $\sum_{i=1}^I n_i^2$ of the charges of the smeared D7/O7 branes is modelled as \[\sum_{i=1}^I n_i^2=m \int_0^\pi d\psi=m\pi\,.\] 
We are lead in this way to:
\begin{equation}
{\cal A}_\pm \Big\lvert_{D7/O7}=-\frac34 \frac{m\pi}{\beta}\Bigl(iw-\frac{\beta\pi}{2}w^2\Bigr),
\end{equation}
where we have also omitted constant terms that are absorbed in ${\cal A}^0_\pm$. This exactly reproduces the linear and quadratic terms proportional to the mass in eq.~\eqref{eq:holo_Ansatz_ATD_new}. The full locally holomorphic functions are then given by ${\cal A}_\pm={\cal A}_\pm |_{NS5}+{\cal A}_\pm |_{D7/O7}$.

We have seen that the Abelian T-dual of the BO solution provides an explicit solution for the annulus with smeared NS5- and D7-branes. The presence of these, however, does not allow for the regularity conditions derived in \cite{DHoker:2017mds,DHoker:2017zwj} to be enforced, which were engineered for localised sources.

\subsection{The AdS$_6$ non-Abelian T-dual solution}
\label{sec:NATD}

The explicit solution constructed in \cite{Lozano:2012au} was obtained dualising the BO solution \eqref{eq:AdS6 sol} with respect to one of the $SU(2)$ isometry groups of the $S^3$. Using the conventions in  \cite{Lozano:2013oma}, it reads
\begin{align}\label{eq:AdS6_sol_NATD}
ds^2&= \frac{L^2 W^2}{4}\bigg[9 ds^2(AdS_6)+ \frac{r^2}{\Delta}\sin^2\theta ds^2(S^2)\bigg]+\frac{4}{L^2W^2\sin^2\theta}\left(dr^2+\frac{1}{4}L^4W^4\sin^2\theta d\theta^2\right) ,\nn\\[2mm]
e^{-\Phi}&=\frac{3L^2}{4W^4}\sin\theta\sqrt{\Delta},~~~B_2=\frac{r^3}{\Delta}\text{Vol}(S^2),~~~F_1=\frac{5L^4}{8 W^2}\sin^3\theta d\theta- m r dr ,\\[2mm]
F_3&=\frac{L^4 r^2\sin^3\theta}{16 \cos\theta W^2 \Delta}\big(-10r \cos\theta d\theta+\sin\theta dr\big)\wedge \text{Vol}(S^2),~~~F_5=0,~~~\Delta=r^2+ \left(\frac{LW\sin\theta }{2}\right)^4. \nn
\end{align}
In this solution $r\in \mathbb{R}^+$, and the internal space is thus non-compact. This is common to all solutions generated with NATD when it is applied on a freely acting $SU(2)$ isometry group, which is replaced by its $\mathbb{R}^3$ Lie algebra after the procedure. Therefore, these solutions need to be completed globally (rendering $r$ a compact direction) in order for them to describe holographically well-defined CFTs\footnote{A non-compact direction of the internal space gives rise to operators with continuous scaling dimensions for the dual CFT.}. We also remark that the $SU(2)$ global symmetry associated to the antisymmetric hypermultiplet of the $USp(2N)$ theory has disappeared in the new solution. This has implications that we will discuss in section \ref{sec:field-theory}.

As the Abelian T-dual, the non-Abelian T-dual solution has two singularities. The  first one is inherited from the D8/O8 system of the BO solution, at $\theta=\pi/2$. The behaviour of the metric close to this singularity is that of D5-branes smeared on $\mathbb{R}^3$, which suggests a smeared D5/O5 system\footnote{This is in agreement with the analysis performed in appendix B of \cite{Lozano:2018pcp}, which shows that worldsheet parity reversal is mapped under non-Abelian T-duality (applied on a freely acting $SU(2)$) onto the combined action of this operation and an inversion in $\mathbb{R}^3$.},  located at $r=0$. However, it is more convenient in our case to think of the O5-fixed plane as an O7 wrapped on a collapsing  $S^2$. The second singularity present in the non-Abelian T-dual solution is at $\theta=0$, where we find NS5-branes smeared on $r$. This is very similar to what we found for the Abelian T-dual solution, since also in this case the $S^3$ along which we dualise shrinks to a point. 

For the computation of the quantised charges, we have to consider that, on each interval $n\pi\leq r<(n+1)\pi$, the $B_2$ field needs to satisfy the constraint  \cite{Lozano:2013oma}, 
\beq
0\leq \frac{1}{4\pi^2} \int_{S^2}B_2 <1\,.
\eeq
This is achieved through a large gauge transformation, $B_2\rightarrow B_2 - n\pi \text{Vol}(S^2)$ of parameter $n$, which runs up to infinity due to the non-compactness of $r$. Taking into account these large gauge transformations, the Page charges in the interval $n\pi\leq r<(n+1)\pi$ read:
\begin{align}\label{eq:Page_charges_NATD}
N_{D7}&=\frac{1}{2\kappa_{10}^2 T_7}\int_{\theta=0}^{\theta=\frac{\pi}{2}} F_1=\frac{9}{32} L^4 m^{1/3} ,\\[2mm]
\label{D5charge}
N_{D5}&=\frac{1}{2\kappa_{10}^2 T_5}\int_{S^2}\int_{\theta=0}^{\theta=\frac{\pi}{2}} (F_3-B_2\wedge F_1)=n N_{D7}, \\[2mm]
\label{ND7r} 
N^r_{7}&=\frac{1}{2\kappa_{10}^2 T_7}\int_{r=n\pi}^{r=(n+1)\pi} F_1=-m\pi^2 (n+\frac12),\\[2mm]
\label{ND5r}
N^r_{5}&=\frac{1}{2\kappa_{10}^2 T_5}\int_{S^2}\int_{r=n\pi}^{r=(n+1)\pi} (F_3-B_2\wedge F_1) = m\pi^2 (\frac{n}{2}+\frac13)  , \\[2mm]
N^n_{NS5}&=\frac{1}{2\kappa_{10}^2 T_5}\int_{S^2}\int_{r=n\pi}^{r=(n+1)\pi} H_3 =1 .
\label{NScharge}
\end{align}
From these, only three are independent charges, as in the Abelian T-dual case. The difference with respect to the Abelian case is that NS5-brane charge keeps being created as we move in $r$, together with D5-brane charge, which is dissolved on D7-branes. This has the effect of duplicating the RR quantised charges associated to the solution. 

It was shown in \cite{Lozano:2016kum,Lozano:2016wrs} that the non-Abelian T-dual solution reduces to the Abelian T-dual one for $r\in[n\pi,(n+1)\pi]$ in the large $n$ limit, where $r$ plays the role of the $\psi\in [0,\pi]$ Abelian T-dual direction. Beyond that limit, and given our lack of knowledge about how D-branes transform under non-Abelian T-duality\footnote{See \cite{Driezen:2018glg} for recent efforts in this direction.}, it is possible to use some arguments based on previous analyses of non-Abelian T-dual solutions \cite{Lozano:2016kum,Lozano:2016wrs,Itsios:2017cew} that support an underlying (D5, NS5, D7/O7) brane system also in the non-Abelian T-dual case. This will be the basis of our analysis in next section when we reproduce the non-Abelian T-dual solution as a DGU geometry.

\subsubsection{The AdS$_6$ non-Abelian T-dual solution as a DGU geometry}

\begin{figure}
	\centering
	\includegraphics[scale=1.3]{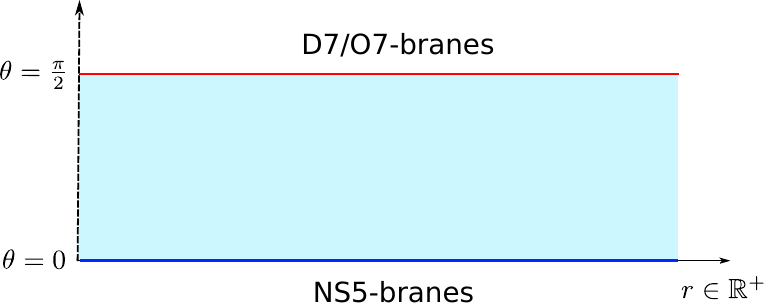}
	\caption{The infinite strip for the non-Abelian T-dual background. NS5-branes are smeared along the lower boundary at $\theta=0$ and D7/O7 branes are smeared along the upper boundary at $\theta=\pi/2$. The strip topology follows from the unboundness of the $r$ direction, $r\in \mathbb{R}^+$.}
	\label{fig:strip_NATD}
\end{figure}

As in the Abelian case, the $(r,\theta)$-directions of the non-Abelian T-dual solution parametrise a Riemann surface $\Sigma$ which, given that $r\in \mathbb{R}^+$, has now the topology of an infinite strip. This is depicted in Figure \ref{fig:strip_NATD}.
Taking the same choice of $w$ as in the Abelian T-dual case,
\begin{equation}\label{eq:NATD_w_def}
w=\frac{1}{\pi}\Bigl( r+\frac{i}{\beta}\left(1-(\cos{\theta})^{2/3}\right)\Bigr),
\end{equation}
the locally holomorphic functions that give rise to the non-Abelian T-dual solution are given by\footnote{These were presented recently in \cite{Hong:2018amk} (using another definition for the $w$ coordinate), where the local embedding in the DGKU classification was considered.}
\begin{equation}
\label{ApmNA}
{\cal A}_\pm=\frac18 m\pi^3 \Bigl(w^3-\frac{3i}{\beta\pi}w^2-\frac{3}{(\beta\pi)^2}w+\frac{i}{(\beta\pi)^3}\Bigr)\pm i\frac34 \pi \Bigl(w-\frac{i}{\beta\pi}\Bigr)+i \frac{m}{4\beta^3}.
\end{equation} 
%As discussed above, we assume a (D5, NS5, D7/O7) brane system similar to the Abelian T-dual case, excepted for the increasing charges computed in \eqref{eq:Page_charges_NATD}.
As discussed above, we have NS5-branes at $\theta=0$, or the real line using \eqref{eq:NATD_w_def}, and D7/O7 branes at $\theta=\pi/2$, corresponding to the upper boundary at ${\rm Im}(w)=\frac{1}{\beta\pi}$.  
Note that in the supergravity limit $\beta\rightarrow 0$ the upper boundary is at infinity. The branes are smeared in $r$, therefore not permitting isolated poles or punctures to arise in ${\cal A}_\pm$. We will now attempt to reproduce ${\cal A}_\pm$ in the $\beta\rightarrow 0$ limit, considering the smearing of the branes, using the upper half-plane formalism in the presence of 7-branes presented in section \ref{sec:upper}.

Let us start analysing the contribution of the NS5-branes, given by the term with $Y_\pm^l$ coefficients in \eqref{Apm7branes}. As discussed above, in the non-Abelian T-dual solution a NS5-brane is created each time $r=n\pi$ is crossed, with this charge being smeared in this direction. Moreover, $r\in \mathbb{R}^+$ implies an infinite number of branes. Therefore, if we want the overall contribution from the different branes to be finite in the $r\rightarrow\infty$ or $n\rightarrow \infty$ limit, we will have to conveniently regularise their charge.

Such a regularisation is found taking NS5-branes to lie at $\frac{r}{\pi}-i\alpha$ for non zero $\alpha$, smearing them in $r\in [0,n\pi]$ and compensating for the charge on the imaginary axis with their images at $\frac{r}{\pi}+i\alpha$. We can then tune $\alpha$ to reproduce the $\pm i\frac34\pi w$ contribution from NS5-branes to ${\cal A}_\pm$ in \eqref{ApmNA}. Taking into account the NS5-brane charge density in the $[0,n\pi]$ interval, given by equation (\ref{NScharge}), we have
\begin{align}
\label{eq:Zlog_approx}
\sum_l Z^l_\pm  \log{(w-p_l)}&= \mp \frac{3}{8}\int_0^{n\pi}\frac{dr}{\pi} \log{\Bigl(\frac{w-\frac{r}{\pi}-i\alpha}{w-\frac{r}{\pi}+i\alpha}\Bigr)} \,.
\end{align}
%where the approximation for made taking a the double limit $n\rightarrow \infty$ and $\alpha\rightarrow \infty$, keeping an adequate finite ratio for $n/\a$.
Considering now that $Y^l_\pm=-Z^l_\pm$ for anti D7-branes, the finite contribution $\pm i\frac34 \pi w$ to ${\cal A}_\pm$ is recovered in the double limit $n\rightarrow \infty$ and $\alpha\rightarrow \infty$, such that $\alpha/n$ is finite and small.

A similar calculation gives rise to the cubic contribution to ${\cal A}_\pm$. In this case we focus on the integral term in \eqref{Apm7branes}. We compute the function $f(z)$, given by eq.~\eqref{fomega}, considering that we have anti D7-branes located at $z_i=\frac{r}{\pi}+\frac{i}{\beta\pi}$, smeared in $r\in [0,n\pi]$, with a charge density given by $dN_{7}^r=-mrdr$, according to \eqref{ND7r}. This yields the following:
%\red{(The charge density $dN_{7}^r=+mrdr$ (positive), with the type of 7-brane $r=-1$, is the right one? We include there both D7 and O7 contributions!)}
\begin{equation}
f(z) 
=-\frac{m}{4\pi} \int_{0}^{n\pi}dr\, r \log{\left(\frac{z-\frac{r}{\pi}-\frac{i}{\beta\pi}}{z-\frac{r}{\pi}+\frac{i}{\beta\pi}}\right)} \,.
%=-\frac{i}{2\pi}m\int_{0}^{n\pi}dr\, r \left[ \frac\pi2 + \arctan{\left(\left(z-\frac{r}{\pi}\right)\beta\pi\right)} \right] .
\end{equation}
We can then reproduce the $\frac18 m\pi^3w^3$ cubic contribution in \eqref{ApmNA} taking the double limit $n\rightarrow \infty$, $\beta\rightarrow 0$ such that $n\beta$ is finite and small, and keeping only finite terms. Adding both contributions of the NS5 and D7/O7 branes, we finally get
\begin{equation}
{\cal A}_\pm ={\cal A}_\pm^0 +\frac18 m\pi^3w^3\pm i\frac34 \pi w.
\end{equation}
These expressions fit the finite terms in $w^3$ and $w$ present in equation \eqref{ApmNA} in the $\beta\rightarrow 0$ limit. However, this regularisation scheme does not allow us to reproduce the infinite, yet $w$-dependent, terms in \eqref{ApmNA}. Unfortunately, a regularisation scheme allowing to completely reproduce \eqref{ApmNA} evaded us.

\subsubsection{On the CFT interpretation of the non-Abelian T-dual solution}
\label{sec:field-theory}

In this section we speculate with a possible 5-brane web compatible with the charges of the non-Abelian T-dual solution. The analysis of charges and background fields performed in section \ref{sec:NATD} suggests that we are dealing with D5-branes extended on $(\mathbb{R}^{1,4},r)$, and NS5-branes extended on AdS$_6$ and localised at $r_p=p\pi$. At each interval $[r_p,r_{p+1})$ there are $pN_{D7}$ D5-branes stretched between the NS5-branes at both ends.  On top of this, there are D7/O7 flavour branes along AdS$_6\times S^2$ whose charges at each interval are given by eq.~\eqref{ND7r}. The brane set-up is symmetric in $x^9$. This is inherited from the $\mathbb{Z}_2$ action of the O8 fixed plane of the BO solution. A summary of this configuration is provided in Table \ref{branes}.
\begin{table}[ht]
	\begin{center}
		\begin{tabular}{| l | c | c | c | c| c | c| c | c| c | c |}
			\hline		    
			& 0 & 1 & 2 & 3 & 4 & 5 & 6 & 7 & 8 & 9 \\ \hline
			D5 & x & x & x & x & x & x &   &   &   &   \\ \hline
			D7/O7 & x & x & x & x & x &   & x & x & x &   \\ \hline
			NS5 & x & x & x & x & x &   &   &   &  & x \\ \hline
		\end{tabular} 
	\end{center}
	\caption{Scheme of the brane intersection proposed for the non-Abelian T-dual solution. In the near-brane limit, $x^9$ would mix with other directions to produce D7/O7 branes spanned on AdS$_6\times S^2$ and NS5-branes spanned on AdS$_6$.}   
	\label{branes}	
\end{table} 

Regarding the contribution of the D7/O7 brane system, from the analysis of the background fields in \eqref{eq:AdS6_sol_NATD} and the transformation of the orientifold action under NATD \cite{Lozano:2018pcp}, it is inferred that in the non-Abelian T-dual solution there is a O5-plane at $r=0$ and $\theta=\pi/2$, which is however smeared in $r$. This O5-plane can be reinterpreted as a O7 wrapped on $S^2$, which shrinks at $r=0$. The non-Abelian T-dual geometry describes only the $r\ge 0$ physical region. In the Abelian T-dual limit, in which $r$ is compactified into a $[0,\pi]$ interval at infinity, there is a second O7 fixed plane at $r=\psi=\pi/2$, which is however missing in the uncompactified case. 

\begin{figure}
	\centering
	\includegraphics[scale=0.3]{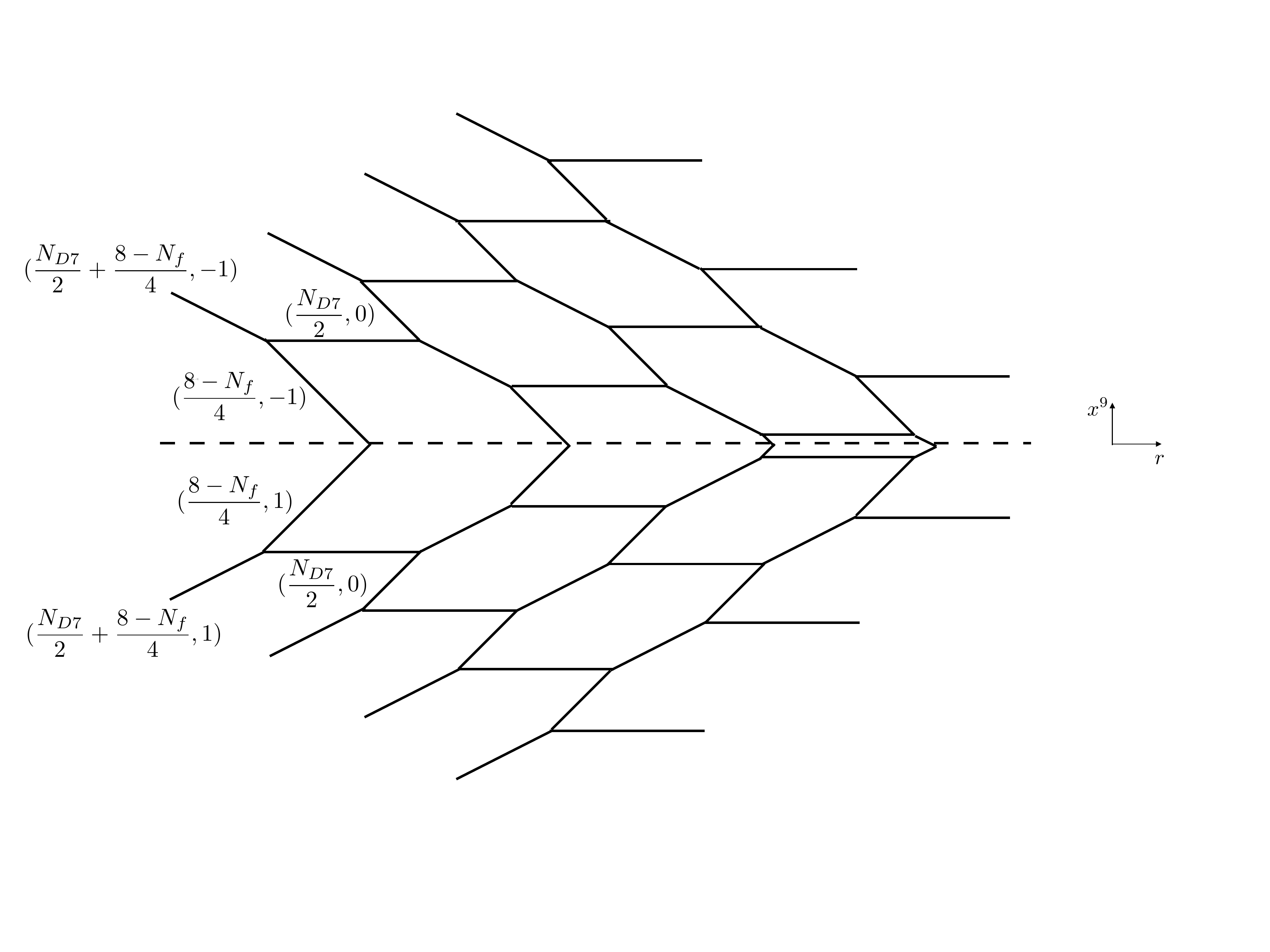}
	\vspace{-1.5cm}
	\caption{5-brane web consistent with the quantised charges of the non-Abelian T-dual solution. The dashed line represents the branch-cut created by the D7/O7 branes at each $r\in [p\pi,(p+1)\pi]$ interval. At each interval a new, smeared, D7/O7 brane system adds to the discontinuity in the 5-brane charge created by the branch cut.}
	\label{extended-set-up}
\end{figure}

Consistently with the previous discussion, we would need to add a 
unique O7 fixed plane \cite{Brunner:1997gk,Bergman:2015dpa} orthogonal to the 5-branes, located at $r=0$, $x^9=0$. The way this can be made compatible with the analysis of the conserved charges is if the fixed plane is smeared in $r$, such that it can give rise,
together with the flavour D7-branes orthogonal to the 5-brane web, to the net $N_7^r=m\pi^2(p+1/2)$ charge at each $r\in [p\pi,(p+1)\pi)$ interval. 
A possible brane set-up compatible with these observations would be the one shown in Figure \ref{extended-set-up}. 

In this brane web the Coulomb branch would correspond to the positions of the D5-branes in the $x^9$ direction. In turn, the lengths of the D5-branes along the $r$-direction would be associated to the inverse of the effective gauge couplings, with the bare couplings arising at the origin of the Coulomb branch. The UV fixed point is then reached by sending these bare couplings to infinity, limit in which the 5-brane web collapses and only the external 5-branes and 7-branes remain.

It is interesting to notice that even if the non-Abelian T-dual solution does not seem to capture many of the global symmetries associated to this brane configuration, the set-up is suggestive of a linear quiver consisting of $USp(2pN)$ gauge groups of increasing rank, as we move from one $[r_p,r_{p+1}]$ interval, with $r_p=p\pi$, to the next in the field theory direction $r\in \mathbb{R}^+$. This interpretation would be consistent with the scaling of the entanglement entropy with the quantised charges, for a spherical entangling surface in DGKU geometries, computed following the prescription provided e.g.~in \cite{Gutperle:2018vdd}. Indeed, this observable scales as $N^{5/2}$ with the colour charge \cite{Lozano:2018pcp}, as expected for $USp(2N)$ gauge theories \cite{Jafferis:2012iv}. 
%We remark, however, that this behaviour seems to be different from the $N^4$ scaling found in \cite{Gutperle:2018vdd} for the $(p,q)$ 5-brane webs considered there. See \cite{Lozano:2018pcp} for more details. 

The previous brane set-up could be seen as a generalisation of the 5-brane webs in the presence of O7 fixed planes describing $USp(2N)$ fixed point theories, discussed in \cite{Bergman:2015dpa}, which satisfy that the number of flavours is $N_F\leq 2N+4$, as required for the existence of a 5d fixed point in the absence of antisymmetric hypermultiplets \cite{Intriligator:1997pq} \footnote{The absence of antisymmetric hypermultiplets is in agreement with the global symmetries of the non-Abelian T-dual solution, as seen in section \ref{sec:NATD}.}. Indeed, this condition can be seen to hold for our set-up, as discussed in \cite{Lozano:2018pcp}.

We should stress that, as in other examples of AdS backgrounds constructed through non-Abelian T-duality \cite{Lozano:2016kum,Lozano:2016wrs,Lozano:2017ole,Itsios:2017cew}, the fact that the field theory direction is non-compact renders an infinite brane set-up, that needs to be regularised such that it can describe a well-defined CFT. Inspired by these examples, we could apply a hard cut-off to the 5-brane web at $r=n\,\pi$, adding an adequate number of D7-branes or external D5-branes so that the brane set-up is consistently terminated. These flavor branes would backreact on the geometry, thus providing a {\it completion} for the background obtained directly from NATD, given in eq.~\eqref{eq:AdS6_sol_NATD}, which would now present an internal space of finite volume. This completed manifold, which is unknown to us, could also realise the global symmetries of the quiver not captured by the non-Abelian T-dual solution. 

%Finally, we would like to stress that, as in the other NATD examples highlighted above, the completion of the field theory, and thereof of the geometry, is not unique. In this example we could for instance have completed the brane set-up by adding a second orientifold fixed plane together with the corresponding mirror image (on $r<0$) of the current brane web. 

\section{Conclusions}
\label{sec:conclusions}

In this proceedings article, we have summarised the main findings in \cite{Lozano:2018pcp}. We have seen that the Abelian and non-Abelian T-duals of the Brandhuber-Oz geometry, the first AdS$_6$ solutions known in Type IIB SUGRA, fit within an extension of the DGU \cite{DHoker:2017mds,DHoker:2017zwj} classification of AdS$_6 \times S^2$ global solutions with 7-branes, that describes smeared 5- and 7-branes. In particular, the explicit realisation of the Abelian T-dual solution required extending the results for the annulus in \cite{DHoker:2017mds} to include 7-branes.  

The T-dual solutions arise from smeared D5, D7/O7 and NS5 branes, that generalise the localised poles and punctures of the solutions in \cite{DHoker:2017mds,DHoker:2017zwj}. These geometries contain non-isolated singularities at the locus of either the D7/O7-system (inherited from the D8/O8 singularity of the original Brandhuber-Oz solution), or the NS5-branes (produced by the T-duality procedure). They terminate the geometry, acting as boundaries that lie infinitely apart in the supergravity limit. The Abelian T-dual solution is associated to a Riemann surface with the topology of an annulus, and the one generated through non-Abelian T-duality is associated to  the upper half-plane.

The Abelian T-dual solution provides the first example of a global solution on the annulus that includes (smeared) 7-branes. 
Apart from recovering this solution, there is an important reason to study the annulus with 7-branes: they may be a requirement for the existence of regular solutions. A numerical analysis on the annulus with just 5-branes was performed in \cite{DHoker:2017mds}, which showed that these solutions cannot satisfy all the regularity conditions. An additional argument for the non-existence of 5-brane solutions on the annulus has been given recently in \cite{Kaidi:2018zkx}. In this reference it is argued  that IIB solutions arising from 5-brane webs with more than one boundary, or higher genus, would lead, upon uplift, to curves in M-theory associated to mass deformations of 5-brane webs, which would correspond to RG flows rather than fixed points. Although this argument seems to contradict the findings of \cite{Lozano:2018pcp}, one should keep in mind that  the M-theory set-up describes a system of localised branes, while the branes present in the Abelian T-dual solution are smeared. Besides, the identification between the Riemann surface and the M-theory curve put forward in \cite{Kaidi:2018zkx} needs not hold for Riemann surfaces with non-trivial topologies. 
It would be interesting to investigate this further -- we also hope to report on solutions with localised 7-branes for the annulus in the future.

 Under certain assumptions, based on the agreement with the Abelian T-dual solution in the large $r$ limit and the similarities with previous examples of non-Abelian T-dual backgrounds \cite{Lozano:2016kum,Lozano:2016wrs,Itsios:2017cew}, a possible 5-brane web with additional D7-branes and a (smeared) O7 fixed plane compatible with the quantised charges of the non-Abelian T-dual solution has been discussed. The associated field theory seems to be consistent with the classification of 5d fixed point theories in \cite{Intriligator:1997pq}.

Our discussion about the possible CFT interpretation of the non-Abelian T-dual solution leaves appealing open problems to study. It would be interesting to investigate in more detail the fixed point theory that arises from a linear quiver of $USp(2N)$ gauge groups with increasing ranks, and its possible relation with the {\it completed} non-Abelian T-dual solution. Further, it would also be striking to explore the global symmetry at the fixed points not visible in supergravity \cite{Bergman:2012kr}, and its possible enhancement, along the lines of \cite{Bergman:2013aca,Bergman:2015dpa}.
The backreaction of the external D7-branes used to regularise the linear quiver  imposing a hard cut-off would provide a geometrical completion of the non-Abelian T-dual solution that could realise these missing global symmetries, along with other properties of the quiver. It should be possible to find out the explicit completed geometry by identifying the locally holomorphic functions associated to the solution with the extra D7-branes.	

%On the other hand, we have discussed that the regularisation procedure yielding a well-defined dual CFT for the non-Abelian T-dual solution is not unique. It would be interesting to provide a consistent framework that singles out one such possible regularisation. For instance in the spirit of \cite{Itsios:2017cew}, that used the connection between 4d $\mathcal{N}=2$ and $\mathcal{N}=1$ CFTs through mass deformations.

\subsection*{Acknowledgements}

We would like to thank the organisers of the Dualities and Generalized Geometries workshop, part of the Corfu Summer Institute 2018 "School and Workshops on Elementary Particle Physics and Gravity", for a very stimulating workshop. Y.L.~and J.M.~are partially supported by the Spanish Government Research Grant FPA2015-63667-P. N.T.M.~is funded by the Italian Ministry of Education, Universities and Research under the Prin project ``Non Perturbative Aspects of Gauge Theories and Strings'' (2015MP2CX4) and INFN.

\bibliographystyle{JHEP}
% The following line makes entries and spaces in between appear smaller:
%\small\baselineskip=.97\baselineskip
\bibliography{AdS6_bibliography.bib}

%\begin{thebibliography}{99}
%\bibitem{...}
%....
%
%\end{thebibliography}

\end{document}